\documentstyle [12pt]{article}
\begin{document}
\def\Arg{\mathop{\rm Arg}}
\def \Journal#1#2#3#4{{#1} {\bf #2}, #3 (#4)}
\def \NCA{\em Nuovo Cimento}
\def \NIM{\em Nucl. Instrum. Methods}
\def \NIMA{{\em Nucl. Instrum. Methods} A}
\def \NPB{{\em Nucl. Phys.} B}
\def \PLA{{\em Phys. Lett.} A}
\def \PLB{{\em Phys. Lett.} B}
\def \PRL{\em Phys. Rev. Lett.}
\def \PREP{{\em Phys. Rep.} C}
\def \PRD{{\em Phys. Rev.} D}
\def \ZPC{{\em Z. Phys.} C}
\def \ApJ{\em Astrophys. J.}
\def \ApJL{\em Astrophys. J. Lett.}
\def \AA{\em Astronomy Astrophys.}
\def \RMPh{\em Rev. Mod. Phys.}
\def \GRG{\em Gen. Rel. Grav.}
\def \CQG{\em Class. Quantum Grav.}
\def \JPA{\em Journal of Physics A}
\def \JETP{\em Sov. Phys. JETP}
\def \ZETP{\em Zh. Exp. Theor. Fiz.}
\def \JP{\em J. Phys. \rm(USSR)}
\def \AP{\em Adv. Phys.}
\def \MNRAS{\em Mon. Not. R. Astron. Soc.}
\def \ASS{\em Astroph. Space Sci.}
\def \PTP{\em Prog. Theoret. Phys.}
\def \PTPS{\em Prog. Theoret. Phys. Suppl. No.}
\def \ha{h}
\def \hdwa{h}
\def \pierwsze{f_1(\eta)}
\def \drugie{f_2(\eta)}
\def \cZero{g_1 }
\def \cJeden{c_1 }
\def \cDwa{c_2 }
\def \cTrzy{g_2 }
\def \czla{\psi_{\lambda}}
\def \czmu{\psi_{\mu}}
\def \czla{f_{\lambda}}
\def \czmu{f_{\mu}}
\def \laIn{\lambda_{n}(\eta)}
\def \muIn{\mu_{n}(\eta)}
\def \laImn{\lambda_{-n}(\eta)}
\def \muImn{\mu_{-n}(\eta)}
\def \lbez{\lambda_{n}}
\def \mbez{\mu_{n}}
\def \n2w{n^2 w}
\def \CLjeden{C_1}
\def \CLdwa{C_2}
\def \bfn{\bf n}
\def \bfx{\bf x}
\def \nx{{\bfn \cdot \bfx}}
\def \ljeden{\log(i\omega\eta)}
\def \ldwa{\log(i\omega\eta)-{1\over \omega^2\eta^2}}
\def \wskaznik{}
\def \widmoA{{\cal A}\wskaznik}
\def \widmoG{{\cal G}\wskaznik}

\title{Acoustics of early universe. II. \\Lifshitz vs. gauge-invariant
theories}
\author{Zdzis{\l}aw A. Golda and Andrzej Woszczyna\\
Astronomical Observatory, Jagellonian University\\
ul. Orla 171, 30--244 Krak\'ow, Poland}
\date{\today}

\maketitle
\begin{abstract}
Appealing to classical methods of order reduction, we reduce the
Lifshitz system to a~second order differential equation. We
demonstrate its equivalence to well known gauge-invariant results. For
a radiation dominated universe we express the metric and density
corrections in their exact forms and discuss their acoustic character.
\end{abstract}


\newpage 

\section{Introduction}

The density perturbations affect the microwave background temperature.
The theory of gravitational instability describes how these inhomogeneities
propagate throughout the radiational era, and foresee the temperature image
they ``paint'' on the last scattering surface. Classical perturbation
theory formulated half a century ago by Lifshitz and Khalatnikov
\cite{lifshitz,lifshitz&khalatnikov,landau&lifshitz}
has nowadays been replaced by more appropriate gauge-invariant descriptions
\cite{olson}--\cite{woszczyna&kulak}. These formalisms introduce some new
measures of inhomogeneity. They do not appeal to the metric tensor, so they
easily avoid spurious perturbations arising from an inappropriate choice of
the equal time hypersurfaces. They guarantee that the space structures
they describe are real physical objects.

On the other hand, the interpretation of the microwave background
temperature fluctuations \cite{cobe} is based on the Sach-Wolfe effect,
where the metric corrections play a key role \cite{sachs&wolfe}.
Therefore, data obtained from COBE is mostly referred to the classical
concepts of Lifshitz and Khalatnikov, and only in a minor part to gauge
invariant measures, which are more precise but difficult to observe
\cite{stoeger&ellis&schmidt}. Both theories in their original
formulations differ essentially. Lifshitz theory provides the two
parameter family of increasing solutions for the density contrast
(\cite{landau&lifshitz} formula (115.19)), while all the gauge-invariant
approaches foresee in concert only a single growing density mode. Thus
the interpretation of the microwave temperature map as the initial data
for cosmic structure formation is fairly ambiguous.

In this paper we attempt to reconcile both types of theories. We appeal
to simple and classical methods of order reduction of differential
equations \cite{matwiejew}. By use of these techniques we remove the
pure-gauge perturbations from Lifshitz theory in the radiation dominated
universe. In consequence we reduce the Lifshitz system to a second order
differential equation, exactly the same as obtained earlier on the
ground of gauge-invariant formalisms. Applying well known solutions,
we express corrections to the metric tensor, the density contrast and
the peculiar velocity in exact form. We show that in the early
universe, scalar perturbations of any length-scale form acoustic waves
propagating with the velocity $1/\sqrt{3}$.

\section{Order reduction}

Relativistic perturbations of a Friedman universe, described in
synchronous coordinates
\cite{lifshitz,lifshitz&khalatnikov,landau&lifshitz} form a
system of two 
second order differential equations with variable coefficients.
In contrast, the similar Newtonian problem is expressed by only
one second-order equation \cite{bonnor,weinberg,padmanabhan}.
Obviously, the two additional degrees of freedom appearing in
the relativistic case must correspond to pure coordinate
transformations (gauge freedom) \cite{lifshitz&khalatnikov}, and
should be removed from the theory.

Removing pure gauge modes we reduce the Lifshitz equations with
pressure $p=\rho/3$ to Bessel equation. The procedure is as follows:
1) we raise the equations order to fourth, in order to separate the
$\muIn$ and $\laIn$ coefficients, and then 2) we reduce the order of
each of the separated equations back by eliminating gauge degrees of
freedom. The resulting equations have exact solutions in the form of
Hankel functions $H_{3/2}$ and their integrals.

In the synchronous system of reference, the metric corrections
$h_{\mu\nu}$ ($\mu,\nu=1,2,3$) to the homogeneous and isotropic,
spatially flat universe fulfill the partial differential equations
\cite{landau&lifshitz} ($8\pi G=c=1$)
\begin{eqnarray}
\ha_{\alpha}^{\phantom{x}\beta ''}
+2{a'\over a}\ha_{\alpha}^{\phantom{x}\beta '}
+(\ha_{\alpha{\phantom{x : y}}:\gamma}^{\phantom{x}\gamma : \beta}
+\ha^{\beta{\phantom{x : y}}:\gamma}_{\phantom{x}\gamma :\alpha}
-\ha_{:\alpha}^{\phantom{: x}:\beta}
-\ha_{\alpha{\phantom{x : y}}:\gamma}^{\phantom{x}\beta :\gamma}
)= 0,\\
2{\left[1+3{dp\over d\epsilon}\right]^{-1}}\left({\ha ''}
+{a'\over a}\left[2+3{dp\over d\epsilon}\right] {\ha '}\right)
+(\ha_{\gamma{\phantom{x : y}}:\delta}^{\phantom{x}\delta : \gamma}
-\ha_{:\gamma}^{\phantom{: x}:\gamma})= 0.
\end{eqnarray}
These equations are usually solved by means of the Fourier transform
\begin{equation}
h_{\mu\nu}= \int{\cal A}({\bf n}) \left[\laIn
\left(\frac{\delta_{\mu\nu}}{3}- \frac{n_{\mu}{n_\nu}} {n^2} \right)
+\frac{1}{3} \muIn
\delta_{\mu\nu}  \right] e^{i\nx}
d^3{\bf n} + c.c.
\label{harm}\end{equation}
The Fourier transform (\ref{harm}) is defined for absolute
integrable functions (the case of least interest for cosmology), for
nonintegrable functions in the framework of distribution theory, or can
be understood as a stochastic integral if the initial conditions are
given at random~\cite{zeldovich&novikov,peebles}. When the
barotropic fluid ($p/\rho =\delta p/\delta \rho = w =\mbox{const}$) is
the matter content of the universe, the functions $\laIn$ and $\muIn$
obey ordinary, second order equations
\begin{eqnarray}
- \n2w (\laIn+\muIn)+2\frac{a'(\eta)}
{a(\eta)}\lbez'(\eta)+\lbez''(\eta)=0,
\label{@L1}\\
- \n2w (1+3w) (\laIn+\muIn)+(2+3w)\frac{a'(\eta)}
{a(\eta)}\mbez'(\eta)+\mbez''(\eta)=0,
\label{@L2}\end{eqnarray}
where prime denotes differentiation with respect to the conformal time
$\eta$ and $a$ is the scale factor for the background metric tensor. In
order to separate the variable $\laIn$, we differentiate (\ref{@L1})
twice and eliminate terms containing $\muIn$ or its derivatives by help
of eq.~(\ref{@L2}). We obtain the fourth order differential equation
\begin{eqnarray}
\left(\n2w\frac{a'(\eta)}{a(\eta)} -
6w\left(\frac{a'(\eta)}{a(\eta)}\right)^3 + 
2 (-1+3w)\frac{a'(\eta)}{a(\eta)} \frac{a''(\eta)}{a(\eta)} + 
2\frac{a^{(3)}(\eta)}{a(\eta)}\right) \lbez'(\eta) \nonumber\\
+\left(\n2w + 6w\left(\frac{a'(\eta)}{a(\eta)}\right)^2 +4\frac{a''(\eta)}{a(\eta)}\right)
\lbez''(\eta) + (4+3 w)\frac{a'(\eta)}{a(\eta)} \lbez^{(3)}(\eta) + \lbez^{(4)}(\eta)=0.
\label{@lam1}\end{eqnarray}
In the same way one can treat (\ref{@L2}) to find the equation for
$\muIn$
\begin{eqnarray} 
\left(\n2w \frac{a'(\eta)}{a(\eta)}-(2+3w)\frac{a'(\eta)}{a(\eta)}
\frac{a''(\eta)}{a(\eta)}+(2+3w)\frac{a^{(3)}(\eta)}{a(\eta)}\right)\mbez(\eta)\nonumber\\
+\left(\n2w+ 2(2+3w) \frac{a''(\eta)}{a(\eta)}\right)\mbez''(\eta)
+(4+3w)\frac{a'(\eta)}{a(\eta)}\mbez^{(3)}(\eta)+\mbez^{(4)}(\eta)=0.
\label{@mu1}\end{eqnarray}

In the following part of this paper we restrict ourselves to a
universe filled with relativistic particles, where both
$w=p_0/\rho_0=\frac{1}{3}$ and ${\cal M}=\rho_0 a^4$ are constants of
motion, and the scale factor $a$ is a linear function of the conformal
time $a(\eta)=\sqrt{{\cal M}/3}\,\eta$. In the flat universe the expansion
rate $\theta(\eta) = 3 a'(\eta)/a(\eta)^2$ and the energy density
$\rho_0(\eta)$ relate to each other by $\rho_0(\eta)=\theta(\eta)^2/3$,
so the equations for $\laIn$ and $\muIn$ take fairly legible form, both
prior to
\begin{eqnarray}
\label{@L1fp}
-\frac{1}{3}{n^2}
(\laIn+\muIn)+\frac{2}{\eta}\lbez'(\eta) +\lbez''(\eta)=0,\\
\frac{2}{3} {n^2}
(\laIn +\muIn) + \frac{3}{\eta}\mbez'(\eta)+\mbez''(\eta)=0,
\label{@L2fp}\end{eqnarray}
and after separation
\begin{eqnarray}
\label{@lam2}
\left(\frac{n^2}{3\eta} - \frac{2}{\eta^3}\right) \lbez'(\eta) +
 \left(\frac{n^2}{3}+\frac{2}{\eta^2}\right)
\laIn+ \frac{5}{\eta}\lbez^{(3)}(\eta)+ \lbez^{(4)}(\eta)=0,\\
\frac{n^2}{3\eta}\mbez'(\eta)
+\frac{n^2}{3}\mbez''(\eta)+ \frac{5}{\eta}\mbez^{3}(\eta) + \mbez^{(4)}(\eta)=0.
\label{@mu2}\end{eqnarray}
We start with equation (\ref{@lam2}). The two well known gauge solutions
\cite{lifshitz} are (with the accuracy to multiplicative constants)
\begin{eqnarray}
\pierwsze &=&1,\\
\drugie &=&-{\sqrt{{\cal M}/3}} \int \frac{1}{a(\eta)}\,d \eta
=-\log(\eta).
\label{fikcyjne}\end{eqnarray}
We expect to obtain  solutions for (\ref{@lam2}) in the
form \cite{matwiejew}: 
\begin{eqnarray}
\laIn={\pierwsze}\left(\int A(\eta)\, d\eta \right),\label{@intA}\\
A(\eta)=\frac{d}{d\eta}\left({\drugie\over\pierwsze}\right)
\left(\int\frac{B(\eta)}{\eta}\,d\eta \right).
\label{@intB}\end{eqnarray}
where $A(\eta)$ and $B(\eta)$ are some auxiliary functions. Inserting
(\ref{@intA}--\ref{@intB}) into (\ref{@lam2}) we obtain the Bessel
equation in its canonical form 
\begin{equation} 
\left(\frac{n^2}{3}-\frac{2}{\eta^2}\right) B(\eta)+ B''(\eta) =0.
\label{@b2}\end{equation}
Equation (\ref{@b2}) is already free of gauge modes, as one can see
from simple heuristic considerations. Let us assume that there exist a
third linearly independent solution of equation (\ref{@L1}), which
corresponds to a pure coordinate transformation. Then, the linear space
of gauge modes would be 3-dimensional, leaving only a single degree of
freedom for the real, physical perturbations. Such a theory has no proper
Newtonian limit.

Equation (\ref{@b2}) is identical to the Sakai equation
(\cite{sakai} formula 5.1), the equation for density perturbations in
orthogonal gauge (\cite{bardeen} formula (4.9), \cite{lyth&mukherjee}
formulae (16--17)), the equation for gauge invariant density gradients
(\cite{ellis&hwang&bruni} formula (38)) or Laplacians (\cite{olson}
formulae (8--9), \cite{woszczyna&kulak} formula (22)) after transforming
these equations to their canonical form (see \cite{golda&woszczyna}). It is interesting to note
that equation (\ref{@b2}) is also identical to the propagation
equation for gravitational waves \cite{grishchuk} (except for
gravitational waves moving with the speed of light). This means that the
solutions to equation (\ref{@b2}) represent waves travelling with the
phase velocity $1/\sqrt{3}$ (we show this explicitly in the next
section). This picture also is consistent\footnote{The procedure we
present here may also be treated as a method to reconstruct metric
corrections and hydrodynamic quantities in their explicit form, out of
the Field and Shepley variables.} with the phonon approach
\cite{lukash}, as the transformation $\phi(\eta)=B(\eta)/\eta+B'(\eta)$
to the Field-Shepley variable
\cite{field&shepley,chibisov&mukhanov} reduces (\ref{@b2}) to
the harmonic 
oscillator $\phi''(\eta)+\frac{n^2}{3}\phi(\eta)=0$ .

\section{Solutions}

The general solution for (\ref{@b2}) is a combination of\footnote{For
similar solutions in the gravitational waves theory see \cite{white}.}
\begin{equation}
B(\eta) = e^{-i\omega\eta}
\left(1+\frac{1}{i\omega\eta}\right) 
\label{B}\end{equation}
and its complex conjugate, with the frequency $\omega =\frac{n} {\sqrt{3}}$. 
These solutions are proportional to Hankel
functions $H_{3/2}$, but more frequently are presented as a combination of
Bessel and Neumann functions $b=a_1 J + a_2 N$ \cite{bardeen}.
Performing integrations
(\ref{@intA}--\ref{@intB}) we determine the solution for $\laIn$ and
find the correction $\muIn$ by solving equation (\ref{@L1})
algebraically

\def\laIn{\lambda(\omega\eta)}
\def \muIn{\mu(\omega\eta)}
\begin{eqnarray}
\laIn &=& -\frac{
e^{-i\omega\eta}}{i\omega\eta}-\mbox{Ei}(-i\omega\eta),
\label{@rlambda1}\\
\muIn &=& \left(1+\frac{1}{i\omega\eta}\right)
\frac{e^{-i\omega\eta}}{i\omega\eta}+\mbox{Ei}(-i\omega\eta).
\label{@rmu1}
\end{eqnarray}
Obviously, equation (\ref{@mu2}) is automatically fulfilled. As a result
we obtain the metric corrections $h_{\mu\nu}$ expanded into planar waves
with the frequency constant in conformal time $\eta$ and with varying
amplitude
\begin{eqnarray}
\hdwa_{\mu\nu}=&-&\int\widmoA({\bf n})\left(\frac{\delta_{\mu\nu}}{3} -
\frac{n_{\mu}{n_\nu}} {n^2} \right)
\left(\frac{e^{i(\nx-\omega\eta)}}{i\omega\eta}+e^{i\nx}\mbox{Ei}(-i\omega\eta)\right)
d^3{\bf n}\nonumber\\
&+&\int \widmoA({\bf n}) \frac{\delta_{\mu\nu}}{3}
\left(\left(1+\frac{1}{i\omega\eta}\right)
\frac{e^{i(\nx-\omega\eta)}}{i\omega\eta}
+e^{i\nx}\mbox{Ei}(-i\omega\eta)\right)d^3{\bf n}+c.c.
\label{h}\end{eqnarray}
The density perturbation and peculiar velocity can be inferred from
formulae (8.2-8.3) of \cite{lifshitz&khalatnikov} and expressed as
\begin{eqnarray}
{\delta\rho\over \rho}\wskaznik&=&
\int\widmoA({\bf n}) u_{\rho}(\nx,\omega\eta) d^3 {\bf n}+c.c.
\label{spanRho}\\
\delta v\wskaznik&=&
\int\widmoA({\bf n}) u_{v}(\nx,\omega\eta) d^3 {\bf n}+c.c.
\label{spanV}\end{eqnarray} 
where the Fourier modes form travelling waves
\begin{eqnarray}
u_{\rho}(\nx,\omega\eta)&=&
{2\over 3}\left(1+\frac{1}{i\omega\eta}+\frac{i\omega\eta}{2}\right)
\frac{e^{i(\nx-\omega\eta)}}{i\omega\eta},\label{ur}\\
u_{v}(\nx,\omega\eta)&=&
{1\over 2\sqrt{3}}\left(1+\frac{i\omega\eta}{2}\right)
\frac{e^{i(\nx-\omega\eta)}}{i\omega\eta}.\label{uv}
\end{eqnarray}
A generic scalar perturbation in the early universe is a superposition
of acoustic waves. Its amplitude decreases to reach a constant and
positive value at late times. This decrease is substantial in the low
frequency (early times) limit $\omega\eta \ll1$. Solutions are formally
divergent at $\eta=0$, nevertheless  evaluating the cosmic structure
backward in time beyond its stochastic initiation $\eta_{i}$ has no well
defined physical sense.

The only perturbations, which are regular at $\eta=0$, and growing near
the initial singularity, consist of standing waves $u({\bf n}\cdot{\bf
x},\omega\eta)+ u({-\bf n}\cdot{\bf x},\omega\eta)$ (compare
\cite{sakai,voglis,peaks} or similar effect in the
gravitational waves theory \cite{grishchuk}). They form a one-parameter
family in the 2-parameter space of all solutions, so they are
non-generic. This property has been confirmed by use of other techniques
in the gauge-invariant theories \cite{ellis&hellaby&matravers}. In the
stochastic approach nongeneric solutions are of marginal interest since
they contribute with the zero probability measure.

\section{Summary and conclusions}

It is a matter of dispute whether cosmic structure was created solely
by gravity forces \cite{lifshitz} or initiated by other,
non-gravitational phenomena manifesting themselves as stochastic
processes \cite{zeldovich&novikov,peebles} in some early epochs.
For the first hypothesis regular and growing solutions are indispensable,
while in the second one the generic perturbations play a key role. In a
radiation dominated universe these properties exclude each other.

Lifshitz theory and the gauge-invariant theories differ less than
usually expected. Both types of theories, when properly written, lead to
the same perturbation equation of the wave-equation form. Generic scalar
perturbations are superpositions of acoustic waves. Solutions depend on
the product $n\eta$ (equivalently on $\omega\eta$). Everything which
concerns early epochs refers also to long waves, and vice
versa\footnote{This is a peculiar property of the spatially flat
radiation-filled universe.}. The perturbation scale does not divide
solutions into different classes. Perturbations propagate with the same
speed $1/\sqrt{3}$, which does not depend on the wave vector. This
confirms the wave nature of scalar perturbations in the radiation
dominated universe (an important property already pointed out by Lukash
\cite{lukash}, but hardly discussed elsewhere) and compels one to use
the complete metric corrections (\ref{h}) in the Sachs-Wolfe procedure
(not only the non-generic growing solutions) at the end of radiational
era.

The reduction technique we apply in this paper can be used for other
equations of state. For $p/\rho =\mbox{const}\neq 1/3$ solutions can be
expressed in terms of hypergeometric functions. In other cases solutions
may not reduce to any known elementary or special functions, although
the reduced equation (\ref{@b2}) can be always found.

\section*{Acknowledgements}

We would like to thank Marek Demia\'nski and Gra\.zyna
Siemieniec-Ozi\c{e}b{\l}o for helpful discussion. This work was
partially supported by State Committee for Scientific Research, project
No 2 P03D 02210.

\section*{Appendix: Lifshitz {`}{`}synchronous{'}{'} gauge}

The original Lifshitz approach
\cite{lifshitz,lifshitz&khalatnikov,landau&lifshitz} provides
solutions 
which are different from (\ref{@rlambda1}--\ref{@rmu1}), and also
inconsistent with the gauge-invariant theories. To explain these
differences in detail, we appeal to the complete solution to
(\ref{@lam2}-\ref{@mu2}) containing both physical and spurious
inhomogeneities. All the gauge freedom within synchronous system is
limited to the choice of the integral constants in (\ref{@intB}).
Actually each of these ``constants" can be defined as an arbitrary
function of the wave number {\bf n} (equivalently $\omega$). We write
them explicitly as ${\cal A}({\bf n})$ and ${\cal G}({\bf n})$
satisfying

\begin{eqnarray}
\laIn={\pierwsze}\left({\cal A}({\bf n})\int A(\eta)\, d\eta -
{\cal G}({\bf n})\log(i\omega)\right)\\
A(\eta)=\frac{d}{d\eta}\left({\drugie\over\pierwsze}\right)
\left(\int\frac{B(\eta)}{\eta}\,d\eta+{{\cal G}({\bf n})
\over {\cal A}({\bf n})} \right).
\end{eqnarray}

so they are equal to the  Fourier coefficients in the
integral
\begin{eqnarray}
\hdwa_{\mu\nu}&=&
\int\widmoA({\bf n})
\left(\frac{n_{\mu}{n_\nu}}{n^2}-\frac{\delta_{\mu\nu}}{3}\right)
\left(\frac{e^{i(\nx-\omega\eta)}}{i\omega\eta}+e^{i\nx}\mbox{Ei}(-i\omega\eta)\right)
d^3{\bf n}
\nonumber\\
&+&\int \widmoA({\bf n}) \frac{\delta_{\mu\nu}}{3}
\left[\left(1+\frac{1}{i\omega\eta}\right)
\frac{e^{i(\nx-\omega\eta)}}{i\omega\eta}
+e^{i\nx}\mbox{Ei}(-i\omega\eta)\right]d^3{\bf n}
\nonumber\\
&+&\int\widmoG({\bf n})
\left[\left(\frac{n_{\mu}{n_\nu}}{n^2}-\frac{\delta_{\mu\nu}}{3}\right)\ljeden
+\frac{\delta_{\mu\nu}}{3}\left(\ldwa\right)\right]e^{i\nx}d^3{\bf n}+c.c.
\label{hg}\end{eqnarray}
Each coefficient ${\cal A}({\bf n})$, ${\cal G}({\bf n})$, can be defined
independently. The gauge freedom is carried by ${\cal G}({\bf n})$  what
follows directly from (\ref{fikcyjne}). Also knowing the gauge invariant
methods one can {\em a posteriori} check that ${\cal A}({\bf n})$
affects the gauge invariant inhomogeneity measures, while ${\cal G}({\bf
n})$ does not. Now, the density contrast and the peculiar velocity, as
inferred from formulae (8.2-8.3) of \cite{lifshitz&khalatnikov}
\begin{eqnarray}
{\delta\rho\over \rho}\wskaznik&=&
\int\left[
\widmoA({\bf n})
u_{\rho}(\nx,\omega\eta)+\widmoG({\bf n}) {\tilde u}_{\rho}(\nx,\omega\eta)
\right]d^3{\bf n}+ c.c.
\label{spanRho1}\\
\delta v\wskaznik&=&
\int\left[
\widmoA({\bf n}) u_{v}(\nx,\omega\eta)+\widmoG({\bf n}) {\tilde u}_{v}(\nx,\omega\eta)
\right]d^3{\bf n}+c.c.
\label{spanV1}\end{eqnarray}
consists of the physical modes $u_{\rho}$, $u_{v}$ already found in
(\ref{ur}-\ref{uv}) and the pure gauge modes equal to
\begin{eqnarray}
{\tilde u}_{\rho}(\nx,\omega\eta)
&=&
{2\over 3}{1\over \eta^2\omega^2}e^{i\nx}\\
{\tilde u}_{v}(\nx,\omega\eta)
&=&
{i\over 2\sqrt{3}}{1\over \omega\eta}e^{i\nx}
\end{eqnarray}
We expand integrals (\ref{spanRho1}) and (\ref{spanV1}) in the early
times limit (with the accuracy to $\eta^2$), to obtain
\begin{eqnarray}
{\delta\rho\over \rho}\wskaznik&=&
\int\left[{2\over 3}{\widmoA({\bf n})+\widmoG({\bf n})\over\omega^2\eta^2}
+\left({1\over 9}i \omega\eta +{1\over 12}\omega^2\eta^2\right)\widmoA({\bf n})
\right]e^{i\nx}d^3{\bf n}+ c.c.\label{spanRho2}\\
\delta v\wskaznik&=&
{1\over 2 \sqrt{3}}\int\left[{\widmoA({\bf n})+\widmoG({\bf n})\over i\omega\eta}
+{1\over 2}\widmoA({\bf n})\left(1+{\omega^2\eta^2\over 6}\right)
\right] e^{i\nx} d^3{\bf n}+ c.c.
\label{spanV2}\end{eqnarray}
Both physical and gauge perturbations manifest identical singular
behaviour at $\eta=0$. Therefore, one cannot distinguish between them
solely on the grounds of their asymptotic forms. On the other hand, one
is able to regularize perturbations by the gauge choice $\widmoG ({\bf
n})=-\widmoA ({\bf n})$. Then, the equal time hypersurfaces follow the
hypersurfaces of equal density at early epochs. This
gauge\footnote{commonly known as the synchronous gauge} has been
actually employed by Lifshitz and Khalatnikov
\cite{lifshitz,lifshitz&khalatnikov}, where divergent terms
${1/\omega^2\eta^2}$ are cancelled by the exactly opposite pure-gauge
corrections\footnote{This does not refer to the metric correction where
${1\over\eta}$-divergence is still present.}. In consequence, perturbations 
described there form a mixture of both the physical and the gauge modes. 

In the Lifshitz gauge, the mode amplitude $ [u_{\rho}({\bf n}\cdot{\bf
x},\omega\eta) \overline{u_{\rho}({\bf n}\cdot{\bf
x},\omega\eta)}]^{1/2}$ grow with time, therefore, the two independent
solutions for the density contrast increase. The same concerns the
peculiar velocity. In the low $\omega\eta$ limit the density contrast
and peculiar velocity form the two-parameter linear spaces of growing
solutions. As a consequence, a generic inhomogeneity increases, which is
in conflict with the gauge-invariant theories
\cite{olson,bardeen,ellis&hwang&bruni}.

\end{document}